\documentclass[article,aps,showpacs,amsfonts,epsf]{revtex4}

\input{epsf.tex}

\newcommand{\E}{{\cal{E}}}
\newcommand{\s}{\sigma}

\renewcommand{\a}{\alpha}
\renewcommand{\k}{\kappa}

\newcommand{\be}{\begin{equation}}
\newcommand{\ee}{\end{equation}}
\newcommand{\bea}{\begin{eqnarray}}
\newcommand{\eea}{\end{eqnarray}}
\newcommand{\ba}{\begin{array}}
\newcommand{\ea}{\end{array}}
\def\J#1#2#3#4{{#1} {\bf #2}, #3 (#4)}
\def\PRD{Phys. Rev. D}
\def\PR{Phys. Rev.}
\def\PRL{Phys. Rev. Lett.}
\def\PTP{Prog. Theor. Phys.}

\def\APL{Ann. Phys. (Leipzig)}

\def\AJ{Astrophys. J.}

\def\JMP{J. Math. Phys.}

\def\CMP{Commun. Math. Phys.}
\def\CQG{Class. Quantum Grav.}

\def\ib{{\it ibid.}}

\begin{document}
\draft
\title{Singularities in the Kerr-Newman and charged $\delta=2$ Tomimatsu-Sato spacetimes endowed with negative mass}

\author{V.~S.~Manko$^\dag$ and E.~Ruiz$^\ddag$}
\address{$^\dag$Departamento de F\'\i sica, Centro de Investigaci\'on y de
Estudios Avanzados del IPN, A.P. 14-740, 07000 M\'exico D.F.,
Mexico\\$^\ddag$Instituto Universitario de F\'{i}sica
Fundamental y Matem\'aticas, Universidad de Salamanca, 37008 Salamanca, Spain}

\begin{abstract}
The Kerr-Newman solution with negative mass is shown to develop a
massless ring singularity off the symmetry axis. The singularity
is located inside the region with closed timelike curves which has
topology of a torus and lies outside the ergoregion. These
characteristics are also shared by the charged Tomimatsu-Sato
$\delta=2$ solution with negative total mass to which in
particular a simple form in terms of four polynomials is provided.
\end{abstract}


\maketitle


\section{Introduction}

In a recent paper \cite{Man} it has been shown that a massless
ring singularity present in the Tomimatsu-Sato $\delta=2$ (TS2)
solution \cite{TSa} and in the Kerr spacetime \cite{Ker} endowed
with negative mass is a locus of points in which the stationary
limit surface (SLS) touches the region with closed timelike curves
(CTC). Since both the TS2 and Kerr solutions are pure vacuum
spacetimes and therefore can not take account of the effects
introduced by the electromagnetic field, it would be interesting
from the physical standpoint to extend the analysis of the paper
\cite{Man} to the charged TS2 \cite{Ern2} and Kerr-Newman (KN)
\cite{NCC} solutions with negative mass in order to find out how
the presence of non-zero charge may affect the location of the
ring singularity with respect to the CTCs region, as well as the
position of the latter region with respect to the SLS. Concerning
the negative-mass KN solution, in the present paper we are going
to establish the most important distinctive feature of that
spacetime, namely, that its ring singularity is {\it massless},
contrary to the massive character of the singularity in the
positive-mass case, the singularity itself being located {\it off}
the SLS and {\it inside} the region with CTCs. We will also
explain why the second asymptotically flat region in the
well-known Carter's maximal extension \cite{Car1} of the KN
spacetime is only capable to give a restricted description of the
negative-mass case. In relation with the charged TS2 solution,
which will be written by us in a new, more concise form than in
the original paper of Ernst \cite{Ern2}, we will show that in the
case when its total mass takes negative values, the location of
the ring singularity and shape of the pathological regions in this
electrovacuum spacetime are qualitatively similar to those of the
KN solution with negative mass.

Our paper is organized as follows. In Sec.~2 we briefly review the
KN solution in the Boyer-Lindquist coordinates and show why the
description of the negative-mass KN geometry in these coordinates
is incomplete. Properties of the KN spacetime endowed with
negative mass are analyzed in the generalized spheroidal and
cylindrical coordinates in Sec.~3 where in particular we obtain
several analytical formulas characterizing the CTCs region and
location of the ring singularity. In Sec.~4 we first rewrite the
charged TS2 solution in a new concise form and then study the case
of the negative total mass. Concluding remarks are given in
Sec.~5.

\section{The KN solution in the Boyer-Lindquist coordinates}

It is well known that the KN solution contains three arbitrary real parameters, $M$, $a$ and $q$, denoting, respectively, the mass, angular momentum per unit mass and charge of the source, and describes the exterior field of a charged rotating black hole when $M$ is positive and satisfies the inequality $M^2\ge a^2+Q^2$. The Ernst complex potentials \cite{Ern} of this solution can be derived straightforwardly from the axis data
\be \E(\rho=0,z)=\frac{z-M-ia}{z+M-ia}, \quad \Phi(\rho=0,z)=\frac{Q}{z+M-ia}, \label{KN_axis} \ee
with the aid of Sibgatullin's integral method \cite{Sib,MSi}, yielding the expressions \cite{MMR}
\be \E=\frac{\k x-M-iay}{\k x+M-iay}, \quad \Phi=\frac{Q}{\k x+M-iay}, \quad \k=\sqrt{M^2-a^2-Q^2}. \label{EF} \ee
The generalized spheroidal coordinates $(x,y)$ employed in (\ref{EF}) are related to the Weyl-Papa\-petrou cylindrical coordinates $(\rho,z)$ by the formulas
\be x=\frac{1}{2\k}(r_++r_-), \quad y=\frac{1}{2\k}(r_+-r_-), \quad r_\pm=\sqrt{\rho^2+(z\pm\k)^2} \label{xy} \ee
($x\ge1$ and $y^2\le1$ for real-valued $\kappa$).

To rewrite $\E$ and $\Phi$ in the ``standard'' Boyer-Lindquist coordinates $(r,\theta)$, one has to perform in (\ref{EF}) the formal substitution
\be \k x=r-M, \quad y=\cos\theta, \label{rt} \ee
thus obtaining
\be \E=\frac{r-ia\cos\theta-2M}{r-ia\cos\theta}, \quad \Phi=\frac{Q}{r-ia\cos\theta}, \label{EF_rt} \ee
with $r\ge0$, $0\le\theta\le\pi$. It follows then that the singularity of the KN solution (zero of the denominator of $\E$) occurs when
\be r=0, \quad \cos\theta=0, \label{sin_rt} \ee
independently of the sign of the mass parameter $M$, which can be also seen from the form of the corresponding Kretschmann scalar ${\cal K}=R_{ijkl}R^{ijkl}$ \cite{Hen}:
\bea {\cal K}&=&\frac{8}{(r^2+a^2\cos^2\theta)^6}[6M^2(r^6-15a^2r^4\cos^2\theta+15a^4r^2\cos^4\theta-a^6\cos^6\theta) \nonumber\\ &&-12MQ^2r(r^4-10a^2r^2\cos^2\theta+5a^4\cos^4\theta) \nonumber\\ &&+Q^4(7r^4-34a^2r^2\cos^2\theta+7a^4\cos^4\theta)]. \label{K}  \eea

However, from (\ref{rt}) one can easily see that whereas the introduction of the Boyer-Lindquist coordinates in the case $M>0$ leads to an extension ($r\ge 0\Rightarrow\k x\ge-M$) of the KN solution (\ref{EF}) and hence is mathematically justified, the transformation (\ref{rt}) in the case $M<0$ causes a contraction ($r\ge 0\Rightarrow\k x\ge|M|$) of the KN manifold, thus suggesting that the Boyer-Lindquist coordinates with $r\ge0$ give a worse description of the negative-mass case then the cylindrical ($\rho,z$) or spheroidal ($x,y$) coordinates.

\begin{figure}[htb]
\centerline{\epsfysize=85mm\epsffile{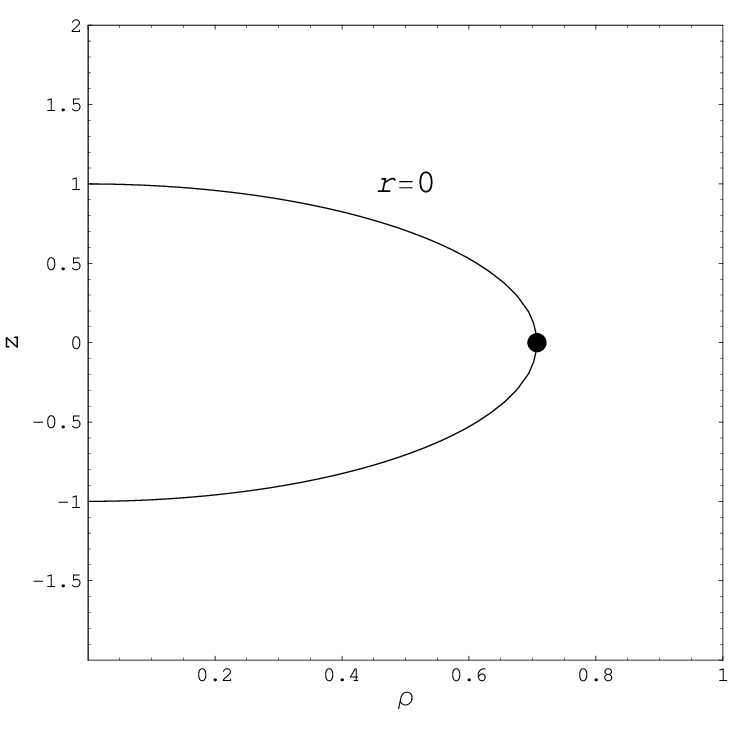}} \caption{The form of the hypersurface $r=0$ in cylindrical coordinates demonstrating that the second asymptotically region of Carter's extension is not enough for the description of the KN spacetime in the $M<0$ case.}
\end{figure}

In this respect it is worth noting that in the classical paper of Carter \cite{Car1} devoted to the maximal analytic extension of the black-hole KN spacetime, the negative-mass case is only briefly mentioned as formally represented by the second asymptotically flat region $M>0$, $r\le0$ of the extended spacetime which is equivalent (via the invariance of the KN metric under the discrete transformation $r\to-r$, $M\to-M$) to the region $M<0$, $r\ge0$. However, as has already been mentioned, the non-negative values of $r$ in the $M<0$ case cover up a more restricted domain than the usual cylindrical coordinates and hence do not provide one with a generic description of the negative-mass case. As an illustration, in Fig.~1 we have plotted the surface $r=0$ in cylindrical coordinated for the particular values of the parameters $M=-1$, $a=Q=1/2$, whence it follows that the spheroid inside the surface $r=0$ is not covered by the range $r\ge0$ of the Boyer-Lindquist coordinates. To cover up the important region represented by the spheroid and containing, as will be seen later on, the ergoregion and the mass source, one needs to extend $r$ to a portion of negative values, as illustrated in Fig.~2 where the region inside $r=0$ is described by $-0.2928<r<0$ (this is equivalent to a redefinition of the radial coordinate $r$ \cite{GMa}). However, such an extension is impossible within the Carter's scheme which employs a specific gluing of two asymptotically flat regions on the surface $r=0$ and has a non-trivial topology. Besides, it is well known that the results of the paper \cite{Car1} are not applicable to the Reissner-Nordstr\"om spacetime \cite{Rei,Nor} which is an electrostatic specialization of the KN solution. For all that, the study of the negative-mass KN case is likely to be performed in a different framework than the specific maximal extension \cite{Car1}.

\begin{figure}[htb]
\centerline{\epsfysize=85mm\epsffile{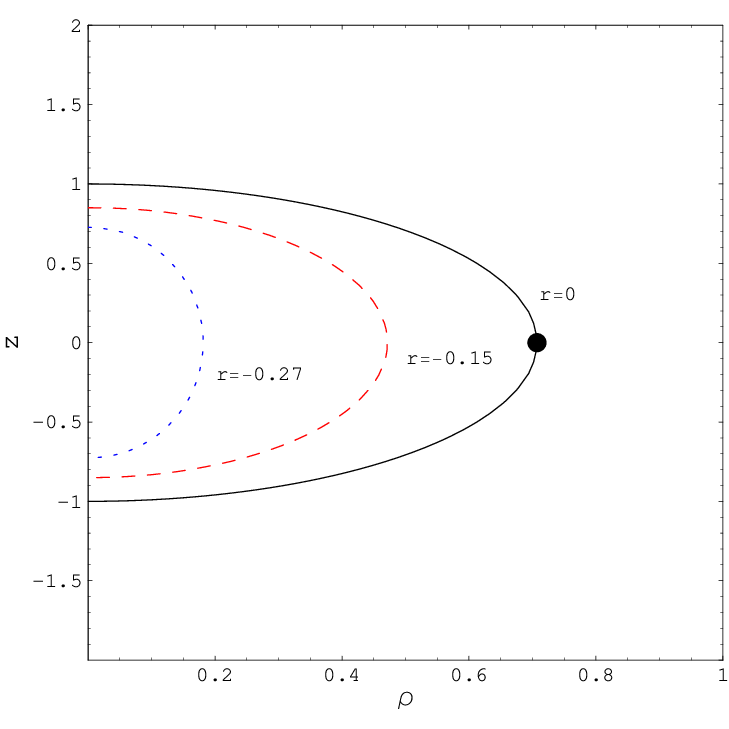}} \caption{A possible (but different from Carter's) extension of $r$ into a restricted range of negative values.}
\end{figure}

\section{The KN metric and its ring singularity in the $M<0$ case}

To analyze conveniently the KN metric in the negative-mass case, we shall consider the corresponding line element in the generalized spheroidal coordinates
\be
d s^2=\k^2f^{-1}\left[e^{2\gamma}(x^2-y^2)\left(\frac{d x^2}{x^2-1}+\frac{d y^2}{1-y^2}\right)+(x^2-1)(1-y^2)d\varphi^2\right]-f(d t-\omega d\varphi)^2, \label{papa} \ee
with the following metric coefficients $f$, $\gamma$ and $\omega$ \cite{MMR}:
\bea f&=&\frac{\kappa^2(x^2-1)-a^2(1-y^2)}{(\kappa
x+M)^2+a^2y^2}, \quad {\rm
e}^{2\gamma}=\frac{\kappa^2(x^2-1)-a^2(1-y^2)}{\kappa
^2(x^2-y^2)}, \nonumber\\ \omega&=&-\frac{a(1-y^2) [2M(\kappa x+M)-Q^2]}{\kappa^2(x^2-1)-a^2(1-y^2)}. \label{mf} \eea

Formulas (\ref{mf}) permit one to study in a unified manner the cases of subextreme (real $\k$) and hyperextreme (pure imaginary $\k$) KN single sources possessing a positive or negative mass. Furthermore, the norm $\eta^\a\eta_\a$ of the axial Killing vector involved in the analysis of the region with CTCs can be shown to have the form
\bea
\eta^\a\eta_\a&=&\kappa^2(x^2-1)(1-y^2)f^{-1}-f\omega^2= \frac{(1-y^2){\cal N}}{(\k x+M)^2+a^2y^2}, \nonumber\\ {\cal N}&=&[(\k x+M)^2+a^2]^2-\k^2a^2(x^2-1)(1-y^2), \label{norm} \eea
while the denominator of the potential $\E$ in (\ref{EF}) gives us the location of the curvature singularity:
\be \k x+M=0, \quad y=0. \label{zero_xy} \ee

In the subextreme black-hole case ($M>0$, $M^2>a^2+Q^2$) extensively discussed by Carter~\cite{Car2} (see also Ref.~\cite{Wal}), extension of the coordinate $x$ to negative values is needed for attaining the singularity because both $M$ and $\k$ in this case are positive definite, as well as $x$ by virtue of (\ref{xy}), while the solution of the first equation in (\ref{zero_xy}) is $x=-M/\k<0$. The introduction of the Boyer-Lindquist coordinates via (\ref{rt}) then implements a standard coordinate extension of the solution leading to (\ref{sin_rt}) instead of (\ref{zero_xy}), the singularity being surrounded by the event horizons $r_\pm=M\pm\k$ (the hypersurfaces $x=\pm1$).

When $M<0$, $M^2>a^2+Q^2$, the first equation in (\ref{zero_xy}) has one positive root
\be x=-\frac{M}{\k}=-\frac{M}{\sqrt{M^2-a^2-Q^2}}>1, \label{root} \ee
and consequently no special extension of coordinates is needed for attaining the corresponding singularity since (\ref{root}) defines, together with $y=0$, a ring located {\it outside} the hypersurface $x=1$ (or $\rho=0$). Indeed, $y=0$ is equivalent to $z=0$ (the equatorial plane), and then (\ref{root}) with the aid of (\ref{xy}) can be trivially solved for $\rho$, yielding instead of (\ref{zero_xy})
\be \rho=\sqrt{a^2+Q^2}, \quad z=0. \label{sin_rz} \ee
Obviously, (\ref{sin_rz}) describes a ring of radius $\sqrt{a^2+Q^2}$ lying in the equatorial plane and having its center at the symmetry axis ($\rho=0$). Mention that formula (\ref{sin_rz}) has been also given independently in a recent paper of Meinel \cite{Mei}.

Remarkably, in the case of hyperextreme or extreme KN sources with negative mass ($M<0$, $M^2\le a^2+Q^2$) the ring singularity is again described by formulas (\ref{sin_rz}), which can be readily verified by rewriting $\k x=-M$, $y=0$ in cylindrical coordinates.

In the absence of rotation ($a=0$) the KN solution reduces to the Reissner-Nordstr\"om spherically symmetric spacetime \cite{Rei,Nor}, and in the case of negative mass, the KN singular ring converts into a singular sphere which in the $\rho$ and $z$ coordinates emerges as a spheroid with the equatorial radius $|Q|$ and the poles at $z=\mp M$. Formally this is due to the fact that the denominator in (\ref{EF}) becomes independent of $y$ when $a=0$, but intrinsically the reason for a new shape of the singularity is a symmetry change. Mention also that in the particular case of the Schwarzschild spacetime with $M<0$, the whole hypersurface $x=1$ becomes singular.

It is not difficult to show that the singularity of the KN solution with $M<0$ is {\it massless}, so that the whole negative mass comes from the segment $x=1$ ($\rho=0$, $|z|\le\k$) of the $z$-axis. Indeed, calculating the Komar mass $M_K$ \cite{Kom} of the latter segment with the aid of Tomimatsu's formula \cite{Tom}
\be M_K=-\frac{1}{4}\omega_0(\Omega|_{x=1,y=1}-\Omega|_{x=1,y=-1}), \label{Tom} \ee
where $\omega_0$ is the value of $\omega$ on the hypersurface $x=1$ and $\Omega$ denotes the imaginary part of the potential $\E$, namely,
\be \Omega=-\frac{2May}{(\k x+M)^2+a^2y^2}, \label{E_im} \ee
one readily arrives at the value $M$ that coincides with the total mass of the KN source obtainable from the asymptotic expansion of the metric coefficient $f$.

Thus, the masslessness of the ring singularity distinguishes
physically the KN spacetimes endowed with negative mass from those
characterized by positive mass \cite{NYa}. Here it is worthy to
note that general relativity is a geometrical theory in which the
singularities are determined via some geometric invariants;
therefore, a ``non-geometric'' question such as whether or not
these singularities may carry certain amount of mass always needs
an additional analysis. In particular, it is easy to see from
(\ref{K}) that in the massless KN case ($M=0$) the ring
singularity is still present and has zero mass, the corresponding
geometry then describing a massless charge superposed with
magnetic dipole moment.

\subsection{Ergosurface and region with CTCs}

The ergosurface, also known as infinite redshift surface, is defined by the equation $f=0$, or explicitly
\be \kappa^2(x^2-1)-a^2(1-y^2)=0, \label{sls} \ee
and its shape does not depend on the sign of the mass parameter
$M$. It is well known that in the subextreme case its topology is
that of a sphere, and in the extreme and hyperextreme cases --
that of a torus. Although in what follows we could restrict our
consideration of this surface to the subextreme case only, the
generalized coordinates ($x,y$) permit us to naturally include
into our analysis the hyperextreme and extreme cases too because
the product $\kappa x$, as it follows from (\ref{xy}), is always
defined as a non-negative function independently of whether
$\kappa$ is real, zero or pure imaginary. Below we will occupy
ourselves with answering the most interesting question about the
ergosurface: what is its location relative to the naked
singularity in the case $M<0$?

To answer this question, we must consider the $S^1$ intersection of the ergosurface with the equatorial plane, i.e., we have to put $y=0$ in (\ref{sls}) and solve the resulting equation for $x$. The positive root then has the form
\be x=\frac{\sqrt{M^2-Q^2}}{\k}, \label{x_sls} \ee
and its value is less than $x=-M/\k$ defining the ring singularity, which means that the singularity is located {\it outside} the ergosurface and ergoregion. In the cylindrical coordinates the above intersection is given by
\be \rho=|a|, \quad z=0, \label{r_sls} \ee
and we see again that this is closer to the symmetry axis than the location of the ring singularity $\rho=\sqrt{a^2+Q^2}$, $z=0$. Only in the absence of charge ($Q=0$) the two rings coincide, the singularity then locating at the equator of the ergosurface.

Another region of interest emerging in the KN spacetime with negative mass is the region with CTCs where the norm of the axial Killing vector (\ref{norm}) takes negative values and causality violation occurs. The boundary of this region is defined by the equation
\be {\cal N}=0, \label{bound} \ee
while the region itself consists of the points for which ${\cal N}<0$.

The analytical study of (\ref{bound}) is more difficult than in the case of the Kerr metric \cite{Ker} with negative mass, and the aggravation is clearly due to the presence of electromagnetic field, but of course a numerical analysis of (\ref{bound}) does nor represent any difficulty. Nonetheless, it is still possible to give analytic proof to the following three general statements ($M<0$, $a\ne0$, $Q\ne0$):
\begin{itemize}
\item[($i$)] the boundary of the region with CTCs has no common points with the ergosurface;
\item[($ii$)] the ring singularity belongs to the region with CTCs;
\item[($iii$)] the region with CTCs lies entirely outside the ergoregion.
\end{itemize}

To prove ($i$), it is sufficient to consider a linear combination of Eq.~(\ref{bound}) with Eq.~(\ref{sls}) multiplied by $a^2(1-y^2)$, thus yielding
\be [(\k x+M)^2+a^2y^2][(\k x+M)^2+a^2(2-y^2)]=0. \label{comb} \ee
The second factor in (\ref{comb}) is always positive definite since $y^2\le1$, while the first factor vanishes at $x=-M/\k$, $y=0$, i.e., at the singularity. The substitution of the latter values into (\ref{sls}) then leads to the condition $Q^2=0$, which contradicts the initial assumption that $Q$ is non-vanishing. This proves ($i$).

To verify ($ii$), it is only necessary to substitute the values of $x$ and $y$ defining the singularity into the expression for ${\cal N}$, thus getting
\be {\cal N}(y=0,x=-M/\k)=-a^2Q^2<0, \label{N_sin} \ee
whence it follows that the ring singularity does belong to the region with CTCs.

Lastly, since the ring singularity, as has already been
established, lies outside the ergoregion and belongs to the region
with CTCs whose boundary has no common points with the
ergosurface, then the latter region lies entirely outside the
ergoregion. Mention that in the hyperextreme case ($M^2<a^2+Q^2$)
the toroidal ergosurface has also the second intersection with the
equatorial plane at $\rho=(a^2+Q^2-M^2)/|a|$, this positive value
being less than $|a|$ because $M^2>Q^2$, and in the extreme case
($M^2=a^2+Q^2$) the latter $\rho$ is equal to zero, so that the
ergosurface touches the symmetry axis at only one point.

The extent in the equatorial plane of the region with CTCs is determined by two real roots of the quartic equation
\be {\cal N}(y=0,X)=[(X+M)^2+a^2]^2-a^2(X^2-\k^2)=0, \quad X:=\k x, \label{N_eq} \ee
which, however, are not given here because of their cumbersome explicit form. Note that in the vacuum limit ($q=0$), Eq.~(\ref{N_eq}) admits partial factorization due to a common ring singularity shared by the ergosurface and boundary of the CTC region and, as a consequence, the corresponding roots have a rather simple form \cite{Man}.

\begin{figure}[htb]
\centerline{\epsfysize=75mm\epsffile{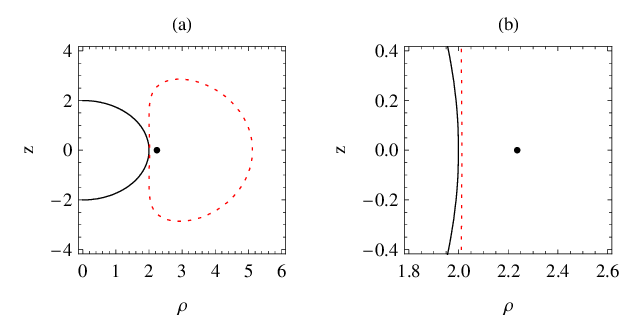}} \caption{In Fig.~3(a) the ergosurface (black curve), region with CTCs (inside the dashed curve) and ring singularity (black dot) of the KN subextreme solution with negative mass are plotted. Fig.~3(b) presents a fragment of the plot 3(a) for illustrating that the ergoregion and region with CTCs do not touch each other.}
\end{figure}

In Figs.~3 and 4 the ergoregion, region with CTCs and ring singularity are plotted for two particular choices of the parameters. Fig.~3(a) represents a typical subextreme case of the KN solution with negative mass ($M=-3$, $a=2$, $Q=1$), and though it might look that the ergosurface touches the boundary of the region with CTCs, this is not really so, as can be well seen from Fig.~3(b), but the two surfaces approach each other very closely. The three positive values of $\rho$ defining the points at which the equatorial plane intersects the ergosurface and boundary of the region with CTCs are the following: 2, 2.013, 5.122, while the location of the ring singularity is defined by $\rho\approx2.236$. Fig.~4 describes the particular hyperextreme case $M=-2$, $a=3$, $Q=1$, for which there are four intersections of the equatorial plane with the ergosurface and boundary of the region with CTCs defined by the positive values of $\rho$: 2, 3, 3.019, 5.167, while the ring singularity is located at $\rho\approx3.162$. Note that whilst in the subextreme case only the region with CTCs has toroidal topology, in the hyperextreme case both the ergoregion and region with CTCs have topology of a torus.

\begin{figure}[htb]
\centerline{\epsfysize=75mm\epsffile{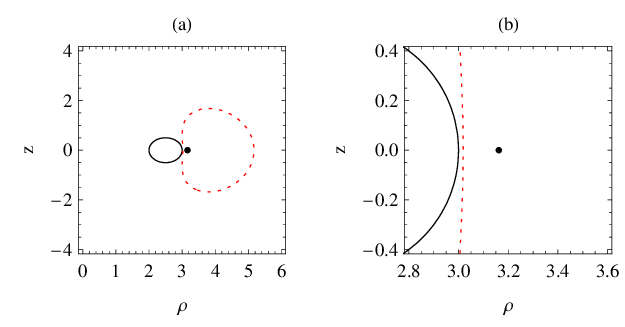}} \caption{In Fig.~4(a) the ergoregion (inside the black curve), region with CTCs (inside the dashed curve) and ring singularity (black dot) of the KN hyperextreme solution with negative mass are plotted. Fig.~4(b) presents a fragment of the plot 4(a) for illustrating that the ergoregion and region with CTCs do not touch each other in the hyperextreme case too.}
\end{figure}

\section{Charged Tomimatsu-Sato $\delta=2$ solution with negative mass}

The general features of the KN solution endowed with negative mass are likely to be shared by other electrovac solutions with negative total mass. To confirm this, in what follows we shall consider a charged version of the well-known TS2 spacetime \cite{TSa} originally obtained by Ernst \cite{Ern2} and later analyzed by Yamazaki \cite{Yam}. This electrovac solution will be presented by us in terms of only four basic polynomials, like this was earlier done in the case of the pure vacuum TS2 metric \cite{Ern3,Per}.

The Ernst potentials of the charged TS2 solution have the form \cite{Ern2}
\bea
\E&=&\frac{(1-b^2)A-(1+b^2)B}{(1-b^2)A+(1+b^2)B}, \quad \Phi=-\frac{2bB}{(1-b^2)A+(1+b^2)B}, \nonumber\\
A&=&p^2(x^4-1)+q^2(y^4-1)-2ipqxy(x^2-y^2), \nonumber\\
B&=&2px(x^2-1)+2iqy(y^2-1), \label{EP_TS} \eea
where the prolate spheroidal coordinates $(x,y)$ are defined as in (\ref{xy}), but without specifying a particular form of the real positive parameter $\k$. The parameters $p$ and $q$ are subject to the constraint $p^2+q^2=1$, while $b$ is the charge parameter. When $b=0$, (\ref{EP_TS}) determines the original TS2 solution \cite{TSa}.

Using the results of the paper \cite{MRS} in which a charging generalization of the Kinnersley-Chitre solution \cite{KCh} was found, it is easy to see (by setting $\a=\beta=Q=0$, $P=1$ in the respective formulas) that the metric functions $f$, $\gamma$ and $\omega$ of the charged TS2 solution can be written in the form
\bea f&=&\frac{N}{D}, \quad e^{2\gamma}=\frac{N}{K_0^2(x^2-y^2)^4}, \quad \omega=-\frac{\k(1-y^2)F}{N}, \nonumber\\
N&=&\mu^2-(x^2-1)(1-y^2)\s^2, \nonumber\\ D&=&N+\mu\pi+(1-y^2)\s\tau, \nonumber\\ F&=&(x^2-1)\s\pi+\mu\tau, \nonumber\\ \mu&=&(1-b^2)[p^2(x^2-1)^2+q^2(1-y^2)^2], \nonumber\\ \s&=&2pq(1-b^2)(x^2-y^2), \nonumber\\ \pi&=&(4/K_0)p^2x[p(1-b^4)(x^2+1)+2(1+b^4)x], \nonumber\\ \tau&=&(4/K_0)pq(y^2-1)[(1-b^4)px+1+b^4], \nonumber\\ K_0&=&p^2(1-b^2). \label{mf_TS} \eea
Remarkably, the norm of the corresponding axial Killing vector can be also expressed in terms of the polynomials $\mu$, $\s$, $\pi$, $\tau$ only, the result being the following concise expression:
\be
\eta_\a\eta^\a=\frac{\k^2(1-y^2)}{D}\{(x^2-1)(\mu+\pi)^2-(1-y^2)[(x^2-1)\s-\tau]^2\}, \label{norm_TS} \ee
which is characteristic of the whole family of charged Kinnersley-Chitre spacetimes \cite{MRS}.

The total mass $M$, angular momentum $J$ and charge $Q$ of the solution (\ref{EP_TS}) are given by the formulas
\be M=\frac{2\k(1+b^2)}{p(1-b^2)}, \quad J=\frac{4\k^2q(1+b^2)}{p^2(1-b^2)}, \quad Q=-\frac{4\k b}{p(1-b^2)}, \label{MJQ} \ee
whence it follows that the positive values of $M$ correspond to $p>0$, $|b|<1$ and $p<0$, $|b|>1$, while $M$ is a negative quantity for $p>0$, $|b|>1$ and $p<0$, $|b|<1$.

\begin{figure}[htb]
\centerline{\epsfysize=85mm\epsffile{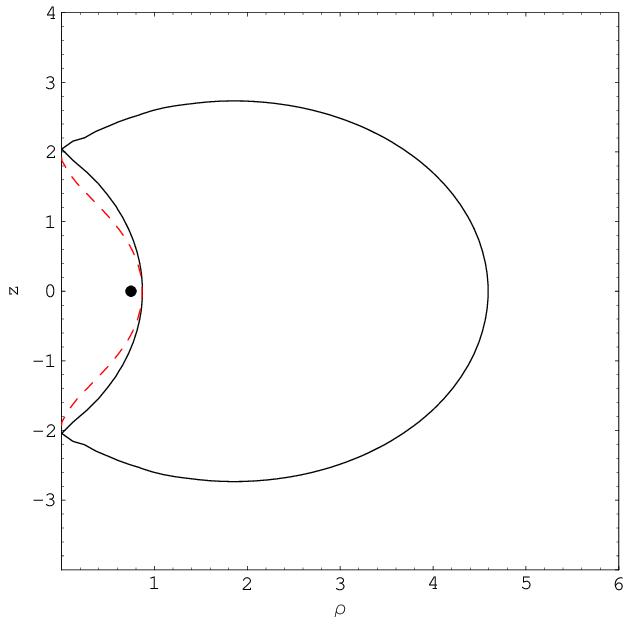}} \caption{The case of the charged TS2 spacetime with $M>0$: the ergoregion (between two solid curves), the region with CTCs (between the $z$-axis and the dashed curve), and the ring singularity (the dot at $\rho\approx0.747$, $z=0$). The particular values assigned to the parameters are $\k=2$, $p=q=1/\sqrt{2}$, $b=1/2$.}
\end{figure}

Before discussing the negative-mass case of the charged TS2 solution, it would be instructive first to briefly consider the positive-mass case of the latter solution with the idea to compare its regions of physical interest with the analogous regions of the vacuum TS2 field, on the one hand, and of the charged TS2 spacetime with $M<0$, on the other hand. In Fig.~5, a typical shape of the SLS, region with CTCs are plotted for the positive-mass charged TS2 spacetime, the ring singularity being located inside the latter region. Compared to a similar diagram \cite{KHi} for the vacuum TS2 solution, Fig.~5 has the following two distinctive features: the ring singularity does not lie on the boundary of the CTC region determined by the dashed line in Fig.~5, and the CTC region does not touch (though comes very close to) the inner SLS at the equator ($y=0$) -- recall that in the uncharged case the ring singularity is a locus of points at the equator in which the boundary of the region with CTCs touches the inner SLS.

\begin{figure}[htb]
\centerline{\epsfysize=65mm\epsffile{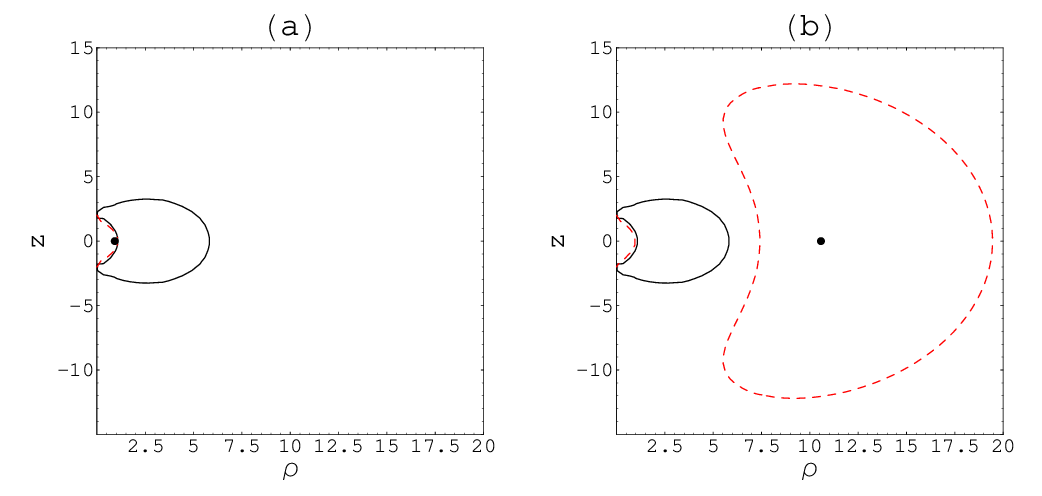}} \caption{The SLSs, ergoregion and ring singularity in the charged TS2 solution with positive (a) and negative (b) total mass. The choice of the parameters is $\k=2$, $p=\pm0.6$, $q=0.8$, $b=0.5$, the singularity being located at $\rho\approx0.924$, $z=0$ (a) and $\rho\approx10.576$, $z=0$ (b).}
\end{figure}

When the total mass in the charged TS2 solution takes negative values, the situation with the problematic region of CTCs and location of the ring singularity becomes qualitatively the same as in the KN solution with negative mass: there appears a large region with CTCs outside the SLS containing the ring singularity inside of it. In Figs.~6(b) and 7(b) this is illustrated with two examples defined by the choices of the parameters $\k=2$, $p=-0.6$, $q=0.8$, $b=0.5$ (Fig.~6(b)) and $\k=2$, $p=0.8$, $q=0.6$, $b=-1.5$ (Fig.~7(b)). Figs.~6(a) and 7(a) represent the positive-mass counterparts of Figs.~6(b) and 7(b), so that the parameter choices in them differ from the above choices for the corresponding figures (b) only in the sign of the parameter $p$. Mention that, similar to the vacuum TS2 field with negative total mass considered in Ref.~1), the CTC region of the charged TS2 solution with $M<0$ consists of two parts, the one inside the inner SLS which is also present in the TS2 solutions with $M>0$, and the second one exterior to the outer SLS.

\begin{figure}[htb]
\centerline{\epsfysize=65mm\epsffile{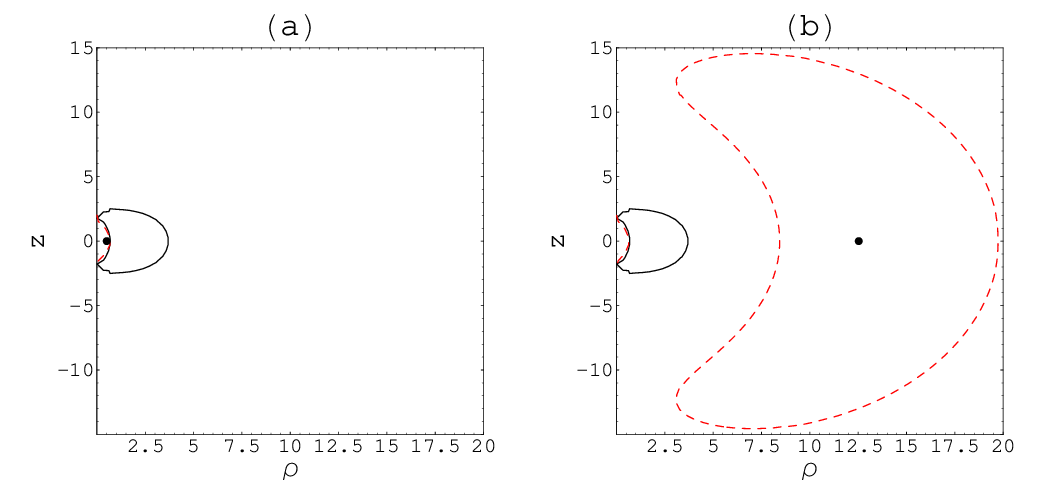}} \caption{The SLSs, ergoregion and ring singularity in the charged TS2 solution with positive (a) and negative (b) total mass. The choice of the parameters is $\k=2$, $p=\mp0.8$, $q=0.6$, $b=-1.5$, the singularity being located at $\rho\approx0.506$, $z=0$ (a) and $\rho\approx12.531$, $z=0$ (b).}
\end{figure}

\section{Concluding remarks}

The present paper may be considered as a useful complement to the known positive mass theorems \cite{SYa,Wit,GHH} for asymptotically flat spacetimes. It clearly demonstrates that negative mass in the KN and charged TS2 solutions is a source of serious pathologies -- a massless ring singularity and region with CTCs, both of which emerge outside the symmetry axis. With respect to the usual KN black-hole geometry where all vicious regions are hidden inside the event horizon and hence are invisible to the exterior observer, the negative-mass KN metric should be considered as a sort of a Wonderland where the exotic physics is exposed for everybody's surprise and inspection. It was suggested by Carter \cite{Car1} that the hidden pathologies in the black-hole case could probably be cured by quantum theory; however, it is not clear how this sort of argumentation could work with regard to massless pathologies of the negative-mass case. In view of the recent work on instabilities of the Schwarzschild solution possessing negative mass \cite{GHI,GDo} it would be plausible to speculate that a possible role of the massless ring singularity in the more general stationary context could be confining negative mass at its location on the symmetry axis, thus ensuring stationarity of the corresponding solution. This may also indicate that the KN spacetime with negative mass is very unlikely to emerge within the gravitational collapse scenarios \cite{Pen,Jos}.

\section*{Acknowledgements}
We are thankful to Malcolm MacCallum for stimulating
correspondence, and to the anonymous referee for valuable
suggestions. This work was partially supported by CONACyT of
Mexico and by MCyT of Spain under the Projects FIS2009-07238 and
FIS2012-30926.


\begin{references}
\bibitem{Man} V.~S.~Manko, \J{\PTP}{127}{1057}{2012}.
\bibitem{TSa} A. Tomimatsu and H. Sato, \J{\PRL}{29}{1344}{1972}; \J{\PTP}{50}{95}{1973}.
\bibitem{Ker} R. P. Kerr, \J{\PRL}{11}{237}{1963}.
\bibitem{Ern2} F. J. Ernst, \J{\PRD}{7}{2520}{1973}.
\bibitem{NCC} E. Newman, E. Couch, K. Chinnapared, A. Exton, A. Prakash and R. Torrence, \J{\JMP}{6}{918}{1965}.
\bibitem{Ern} F.~J.~Ernst, \J{\PR}{168}{1415}{1968}.
\bibitem{Car1} B.~Carter, \J{\PR}{174}{1559}{1968}.
\bibitem{Sib} N.~R.~Sibgatullin, {\it Oscillations and Waves in
Strong Gra\-vitational and Electromagnetic Fields} (Nauka, Moscow,
1984) [English translation (Springer-Verlag, Berlin, 1991)].
\bibitem{MSi} V.~S.~Manko and N.~R.~Sibgatullin, \J{\CQG}{10}{1383}{1993}.

\bibitem{MMR} V. S. Manko, J. Mart\'in and E. Ruiz, \J{\CQG}{23}{4473}{2006}.
\bibitem{Hen} R. C. Henry,  \J{\AJ}{535}{350}{2000}.
\bibitem{GMa} H. Garc\'ia-Compe\'an and V. S. Manko, ArXiv:1205.5848, v.4 (2013).

\bibitem{Rei} H.~Reissner, \J{\APL}{355}{106}{1916}.
\bibitem{Nor} G.~Nordstr\"om, Proc. K. Ned. Akad. Wet. {\bf 20}, 1238 (1918).

\bibitem{Car2} B. Carter, in {\it Black Holes}, edited by C. DeWitt and B. S. DeWitt (Gordon and Breach, New York, 1973), p. 57.
\bibitem{Wal} R. M. Wald, {\it General Relativity} (Chicago: The University of Chicago Press, 1984).
\bibitem{Mei} R.~Meinel, \J{\CQG}{29}{035004}{2012}.
\bibitem{Kom} A.~Komar, \J{\PR}{113}{934}{1959}.
\bibitem{Tom} A. Tomimatsu, \J{\PTP}{72}{73}{1984}.
\bibitem{NYa} E. T. Newman and A. I. Janis, \J{\JMP}{6}{915}{1965}.

\bibitem{Yam} M. Yamazaki, \J{\JMP}{19}{1376}{1978}.
\bibitem{Ern3} F. J. Ernst, \J{\JMP}{17}{1376}{1976}
\bibitem{Per} Z. Perj\'es, \J{\JMP}{30}{2197}{1989}.
\bibitem{MRS} V.~S.~Manko, E. Ruiz and M. B. Sadovnikova, \PRD{84,2011,064005}.
\bibitem{KCh} W. Kinnersley and D. M. Chitre, \J{\JMP}{19}{2037}{1978}.
\bibitem{KHi} H. Kodama and W. Hikida, \J{\CQG}{20}{5121}{2003}.

\bibitem{SYa} R. Schoen and S.-T. Yau, \J{\CMP}{65}{45}{1979}; \J{\ib}{79}{231}{1981}.
\bibitem{Wit} E. Witten, \J{\CMP}{80}{381}{1981}.
\bibitem{GHH} G. W. Gibbons, S. W. Hawking, G. T. Horowitz and M. J. Perry, \J{\CMP}{88}{295}{1983}.
\bibitem{GHI} G. W. Gibbons, S. A. Hartnoll and A. Ishibashi, \J{\PTP}{113}{963}{2005}.
\bibitem{GDo} R. J. Gleiser and G. Dotti, \J{\CQG}{23}{5063}{2006}.
\bibitem{Pen} R.~Penrose, \J{\PRL}{14}{57}{1965}.
\bibitem{Jos} P. S. Joshi, {\it Gravitational Collapse and Spacetime Singularities} (Cambridge: Cambridge University Press, 2007).


\end{references}
\end{document}